\documentclass{aip-cp}

\usepackage[numbers]{natbib}
\usepackage{rotating}
\usepackage{graphicx}

\begin{document}

\title{Progress in Neutron EM Couplings}

\author[aff1]{Igor Strakovsky\corref{cor1}}
\author[aff1]{William Briscoe}
\author[aff2,aff1]{Alexander Kudryavtsev}
\author[aff2]{Viacheslav Kulikov}
\author[aff2]{Maxim Martemianov}
\author[aff2]{Vladimir Tarasov}
\author[aff1]{Ron Workman}

\affil[aff1]{Department of Physics, The George Washington 
	University, Washington, D.C. 20052, USA}
\affil[aff2]{Institute of Theoretical and Experimental 
	Physics, SRC ``Kurchatov Institute", Moscow
	117218, Russia}
\corresp[cor1]{Corresponding author: igor@gwu.edu}
\maketitle

\begin{abstract}
An overview of the GW SAID and ITEP groups' effort to 
analyze pion photoproduction on the neutron-target will 
be given. The disentanglement of the isoscalar and 
isovector EM couplings of $\rm N^\ast$ and $\rm 
\Delta^\ast$ resonances does require compatible data on 
both proton and neutron targets. The final-state 
interaction plays a critical role in the state-of-the-art 
analysis in extraction of the $\rm \gamma n\to\pi N$ data 
from the deuteron target experiments. Then resonance 
couplings determined by the SAID PWA technique are then 
compared to previous findings. The neutron program is 
important component of the current JLab, MAMI-C, SPring-8, 
ELSA, and ELPH studies.
\end{abstract}

\section{Introduction}

The $\rm N^\ast$ family of nucleon resonances has many well
established members~\cite{PDG}, several of which exhibit
overlapping resonances with very similar masses and 
widths but with different $\rm J^{\rm P}$ spin-parity values. 
Apart from the $\rm N(1535)1/2^-$ state, the known proton 
and neutron photo-decay amplitudes have been determined 
from analyses of single-pion photoproduction. There are 
two closely spaced states above $\rm \Delta$(1232)$3/2^+$: 
$\rm N(1520)3/2^-$ and $\rm N(1535)1/2^-$. Up to W $\sim$ 
1800~MeV, this region also encompasses a sequence of six 
overlapping states: $\rm N(1650)1/2^-$, $\rm N(1675)5/2^-$, 
$\rm N(1680)5/2^+$, $\rm N(1700)3/2^-$, $\rm N(1710)1/2^+$, 
and $\rm N(1720)3/2^+$. The present work reviews the region 
from the threshold to the upper limit of the SAID analyses, 
which is CM energy W = 2.5~GeV. 

One critical issue in the study of meson photoproduction
on the nucleon comes from isospin. While isospin can 
change at the photon vertex, it must be conserved at the 
final hadronic vertex. Only with good data on both proton 
and neutron targets can one hope to disentangle the 
isoscalar and isovector electromagnetic (EM) couplings 
of the various $\rm N^\ast$ and $\rm \Delta^\ast$ 
resonances (see Refs.~\cite{ref2}), as well as the 
isospin properties of the non-resonant background 
amplitudes. The lack of $\rm \gamma n\rightarrow\pi^-p$ 
and $\rm \gamma n\rightarrow\pi^0n$ data does not allow 
us to be as confident about the determination of neutron 
EM couplings relative to those of the proton. For 
instance, the uncertainties of neutral EM couplings of 
$\rm 4^\ast$ low-lying $\rm N^\ast$ resonances, $\rm 
\Delta(nA_{1/2})$ vary between 25 and 140\% while 
charged EM couplings, $\rm \Delta(pA_{1/2})$, vary 
between 7 and 42\%. Some of the $\rm N^\ast$ baryons 
[$\rm N(1675)5/2^-$, for instance] have stronger EM 
couplings to the neutron relative to the proton, but 
the parameters are very uncertain~\cite{PDG}. One more 
unresolved issue relates to the second $\rm P_{11}$, 
$\rm N(1710)1/2^+$. That is not seen in the recent 
$\rm\pi N$ partial-wave analysis (PWA)~\cite{ref4}, 
contrary to other PWAs used by the PDG14~\cite{PDG}. A 
recent brief review of its status is given in 
Ref.~\cite{ref5}.

Additionally, incoherent pion photoproduction on the 
deuteron is interesting in various aspects of nuclear 
physics, and particularly, provides information on the 
elementary reaction on the neutron, i.e., $\rm \gamma 
n\rightarrow\pi N$. Final-state interaction (FSI) 
plays a critical role in the state-of-the-art analysis 
of the $\rm \gamma n\rightarrow\pi N$ interaction as 
extracted from $\rm \gamma d\rightarrow\pi NN$ 
measurements. The FSI was first considered in 
Refs.~\cite{ref6} as responsible for the near-threshold 
enhancement (Migdal-Watson effect) in the NN mass 
spectrum of the meson production reaction $\rm 
NN\rightarrow NNx$. In Ref.~\cite{ref8}, the FSI 
amplitude was studied in detail.

\section{Complete Experiment in Pion Photoproduction}

Originally, PWA arose as the technology to determine
amplitude of the reaction via fitting scattering data.
That is a non-trivial mathematical problem -- looking 
for a solution of ill-posed problem following to 
Hadamard and Tikhonov~\cite{hati}.  Resonances appeared 
as a by-product (bound states objects with definite 
quantum numbers, mass, lifetime and so on).

There are 4 independent invariant amplitudes for a 
single pion photoproduction. In order to determine 
the pion photoproduction amplitude, one has to carry 
out 8 independent measurements at fixed (s, t) (the 
extra observable is necessary to eliminate a sign 
ambiguity).

There are 16 non-redundant observables and they are 
not completely independent from each other, namely 
1 unpolarized, $\rm d\sigma/d\Omega$; 3 single 
polarized, $\Sigma$, T, and P; 12 double polarized, 
E, F, G, H, $\rm C_x$, $\rm C_z$, $\rm O_x$, $\rm 
O_z$, $\rm L_x$, $\rm L_z$, $\rm T_x$, and $\rm T_z$ 
measurements.  Additionally, there are 18 
triple-polarization asymmetries [9 (9) for linear 
(circular) polarized beam and 13 of them are 
non-vanishing]~\cite{ref10}. Obviously, the 
triple-polarization experiments are not really 
necessary from the theoretical point of view while 
such measurements will play a critical role to keep 
systematics under control.

\section{Neutron Database}

Experimental data for neutron-target photoreactions 
are much less abundant than those utilizing a proton 
target, constituting only about 15\% of the present 
worldwide known GW SAID database~\cite{ref12}. The 
existing $\rm \gamma n \rightarrow\pi^-p$ database 
contains mainly differential cross sections and 15\% 
of which are from polarized measurements. At low to 
intermediate energies, this lack of neutron-target 
data is partially compensated by experiments using 
pion beams, e.g., $\rm \pi^-p\rightarrow\gamma n$, 
as has been measured, for example, by the Crystal 
Ball Collaboration at BNL~\cite{ref13} for the 
inverse photon energy E = 285 -- 689~MeV and $\rm 
\theta = 41^{\circ} - 148^{\circ}$, where $\rm 
\theta$ is the inverse production angle of $\rm \pi^-$ 
in the CM frame.  This process is free from 
complications associated with the deuteron target. 
However, the disadvantage of using the reaction 
$\rm \pi^-p\rightarrow\gamma n$ is the 5 to 500 
times larger cross sections for $\rm \pi^-p
\rightarrow\pi^0n\rightarrow\gamma\gamma n$, 
depending on E and $\rm \theta$, which causes a 
large background, and there were no ``tagging" 
high flux pion beams.
\begin{figure}[htb!]
\centerline{
\includegraphics[scale=0.14,angle=90]{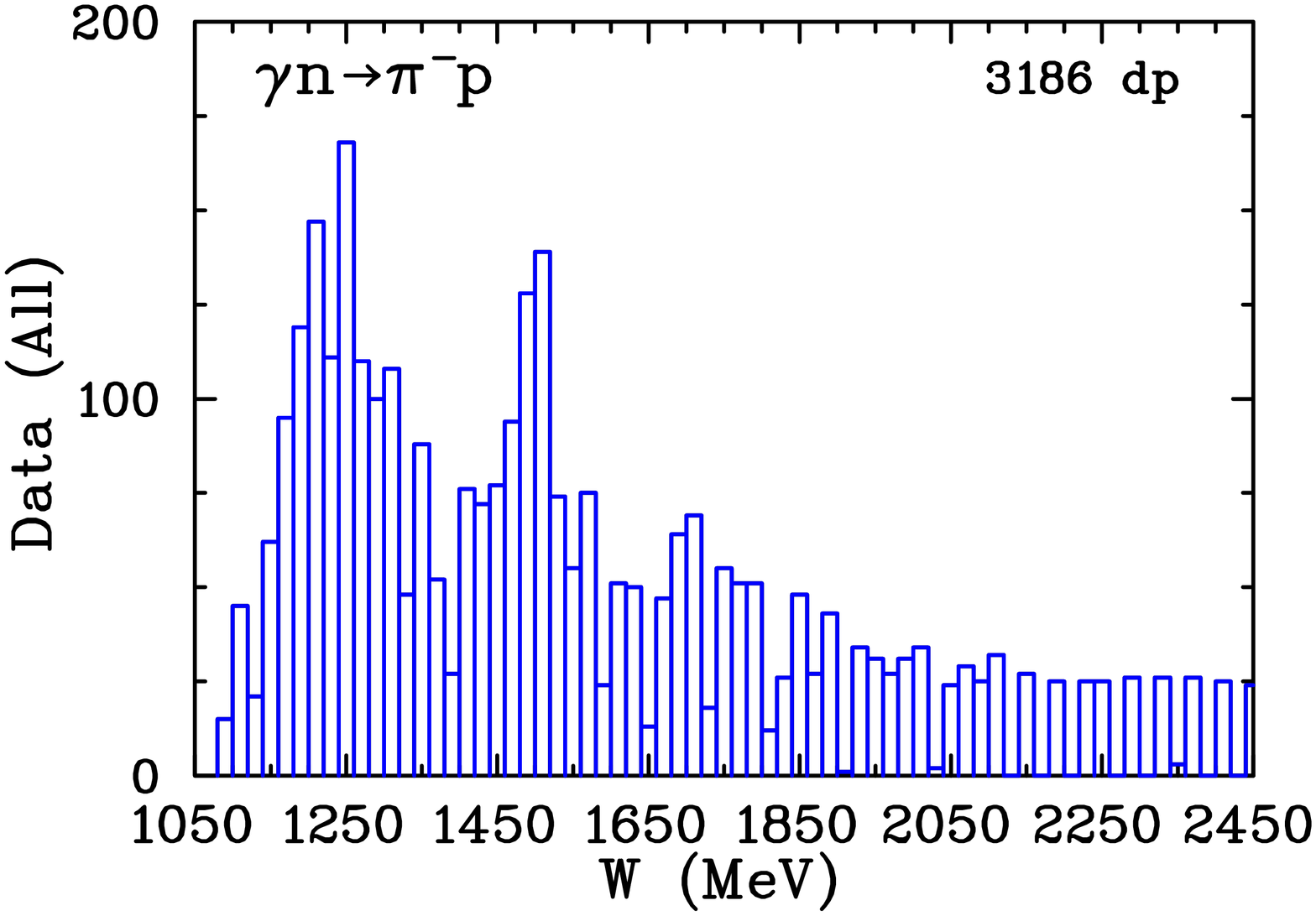}~~~~~~
\includegraphics[scale=0.14,angle=90]{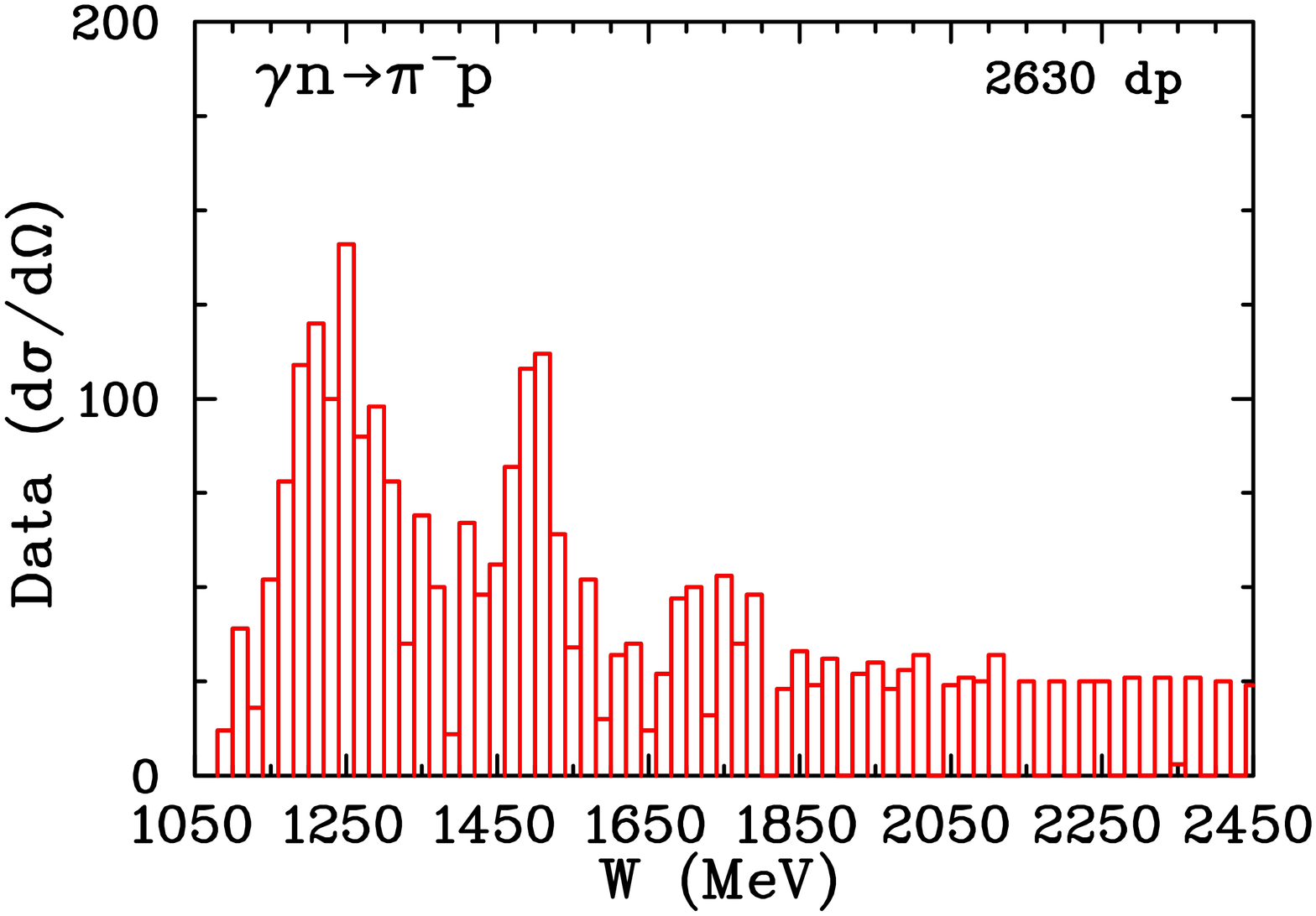}~~~~~~
\includegraphics[scale=0.14,angle=90]{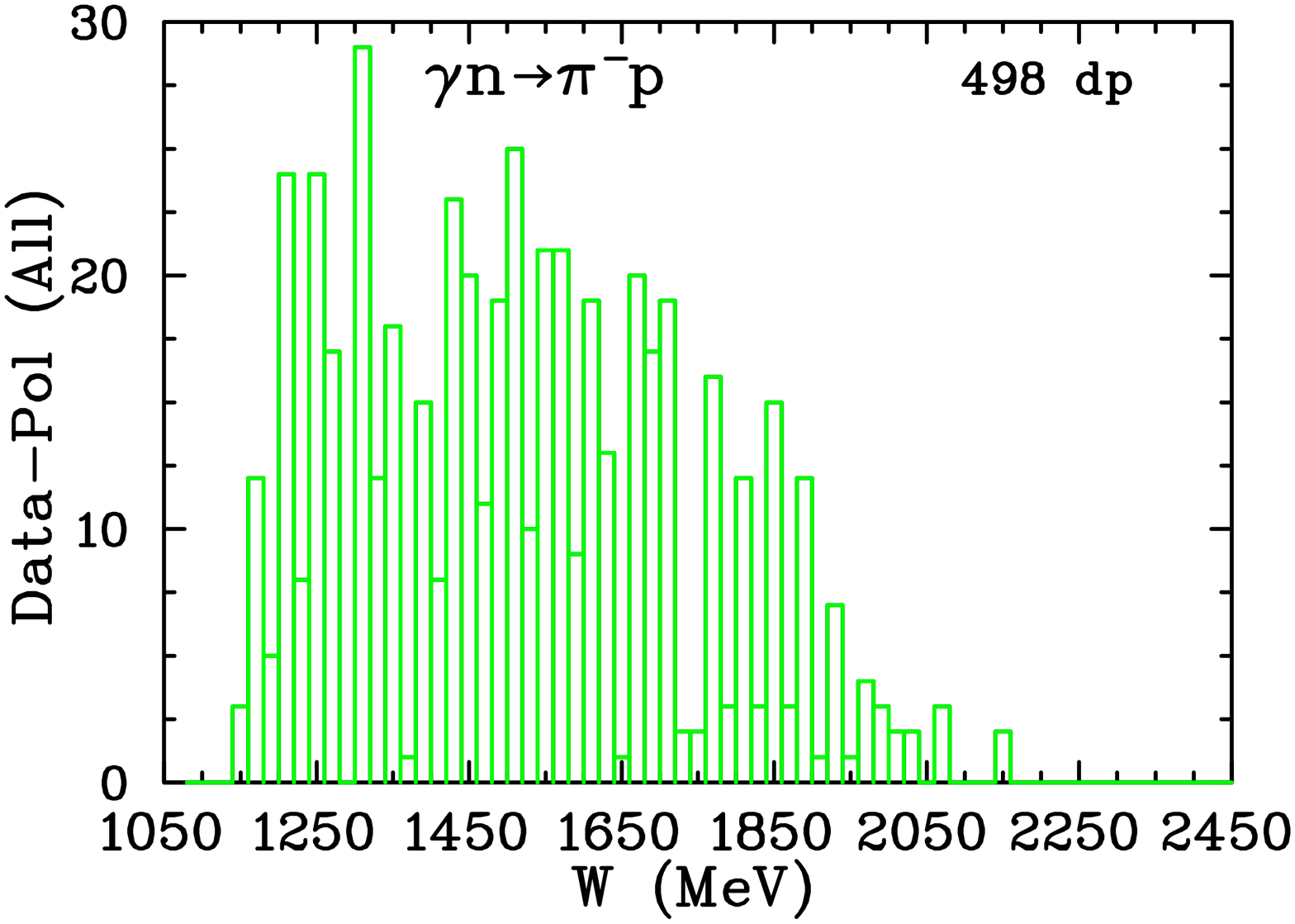}}

\caption{\small{Data available for $\rm \gamma 
	n\to\pi^-p$ as a function of CM energy 
	W~\protect\cite{ref12}.
        The number of data points, dp, is given in 
	the upper right hand side of each subplot.
        The first subplot (blue) 
	shows the total amount of $\rm \gamma n
	\rightarrow\pi^-p$ data available for all 
	observables, the second subplot (red) shows 
	the amount of $\rm d\sigma/d\Omega$ data 
	available, the third subplot (green) shows 
	the amount of polarization observables P
	data available.}} 
	\label{fig:fig1}
\end{figure}
\begin{figure}[htb!]
\centerline{
\includegraphics[scale=0.14,angle=90]{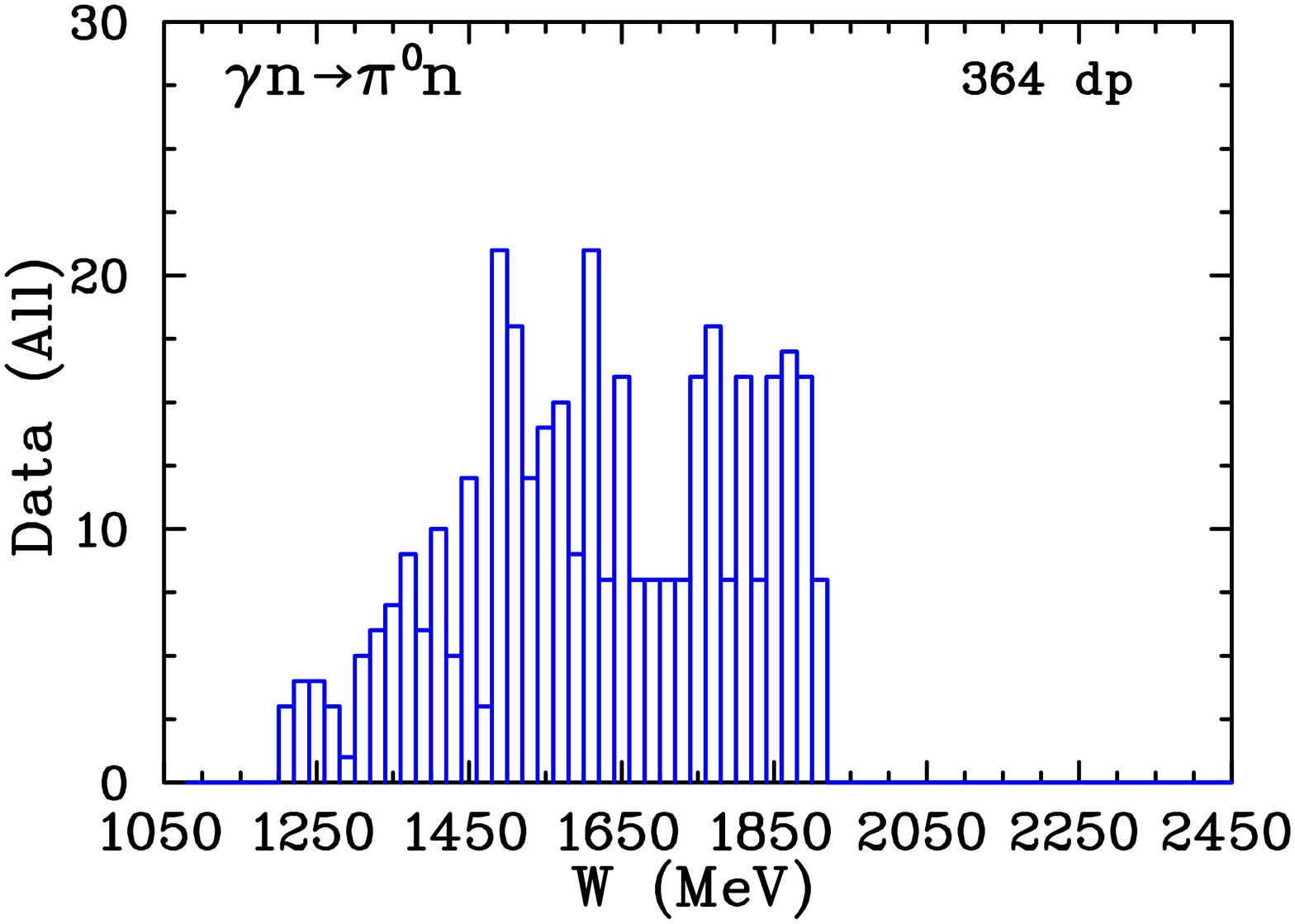}~~~~~
\includegraphics[scale=0.14,angle=90]{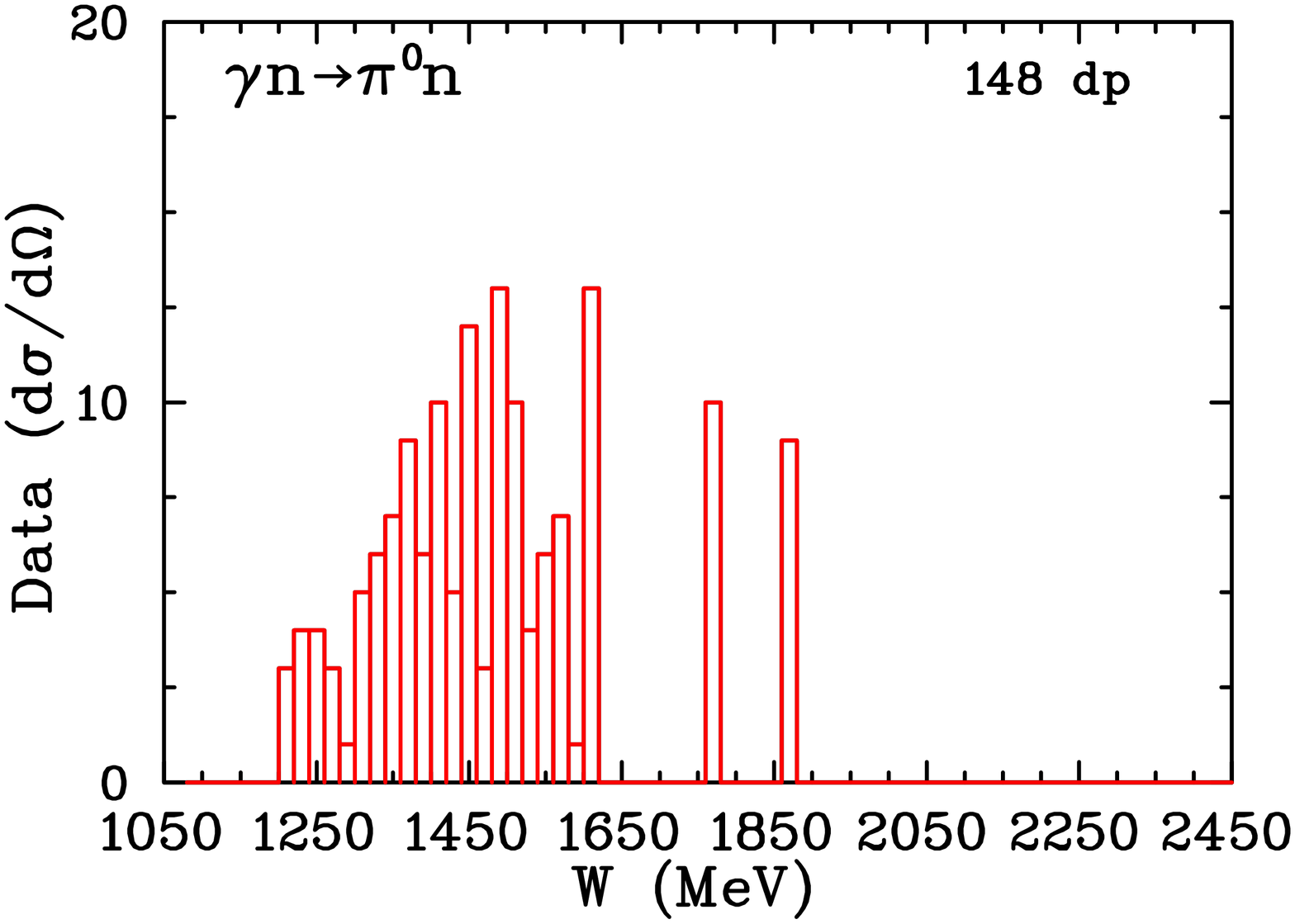}~~~~~
\includegraphics[scale=0.14,angle=90]{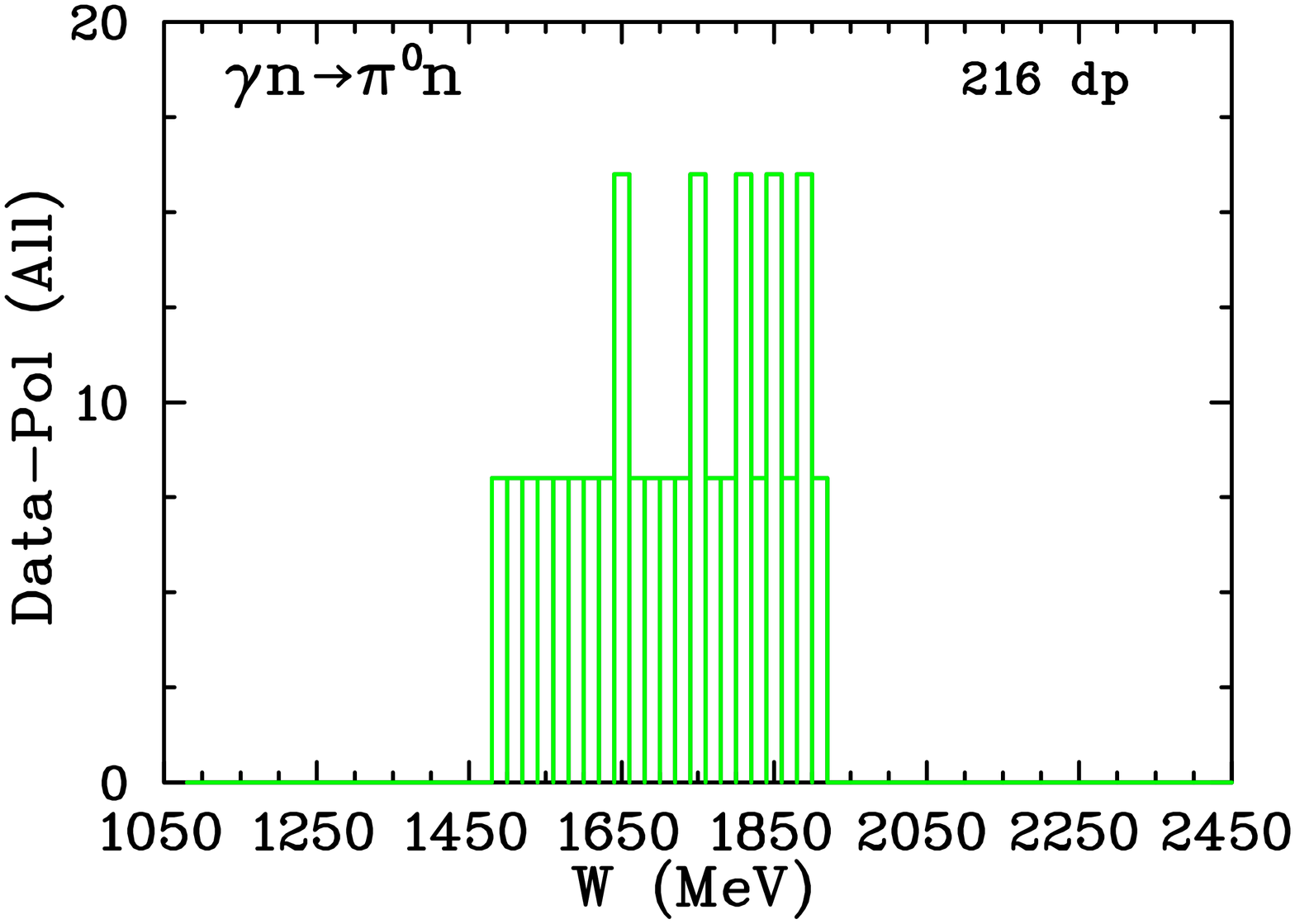}}

\caption{\small{Data available for $\rm \gamma 
	n\to\pi^0n$ as a function of CM energy
        W~\protect\cite{ref12}. Notation as in 
	Fig.~\protect\ref{fig:fig1}.}}
        \label{fig:fig2}
\end{figure}
Figures~\ref{fig:fig1} and \ref{fig:fig2} summarize
the available data for single pion photoproduction on 
the neutron below W = 2.5~GeV. Some high-precision 
data for the $\rm \gamma n\rightarrow\pi^-p$ and 
$\rm \gamma n\rightarrow\pi^0 n$ reactions have 
been measured recently. We applied our GW-ITEP FSI 
corrections, covering a broad energy range up to E = 
2.7~GeV~\cite{ref8}, to the CLAS Collaboration
(E = 1050 -- 2700~MeV and $\rm \theta = 32^{\circ} - 
157^{\circ}$)~\cite{ref14} and A2 Collaboration at 
MAMI (E = 301 -- 455~MeV and $\rm \theta = 45^{\circ} -
125^{\circ}$)~\cite{ref15} $\rm \gamma d\rightarrow
\pi^-pp$ measurements to get elementary cross sections 
for $\rm \gamma n\rightarrow\pi^-p$. In particular, 
the new CLAS cross sections have quadrupled the world 
database for $\rm \gamma n\rightarrow\pi^-p$ above 
E = 1~GeV. The FSI correction factor for the CLAS 
and A2 kinematics was found to be small, $\rm 
\Delta\sigma/\sigma < 10$\%.
\begin{figure}[htb!]
\centerline{
\includegraphics[scale=0.53,angle=0]{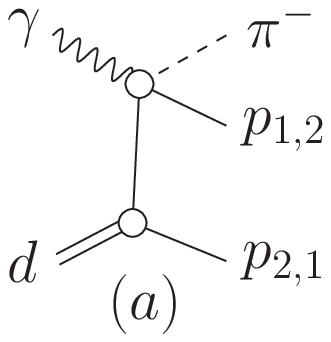}~~~~~~~~
\includegraphics[scale=0.53,angle=0]{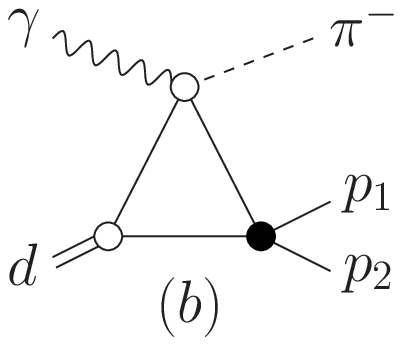}~~~~~~~~
\includegraphics[scale=0.53,angle=0]{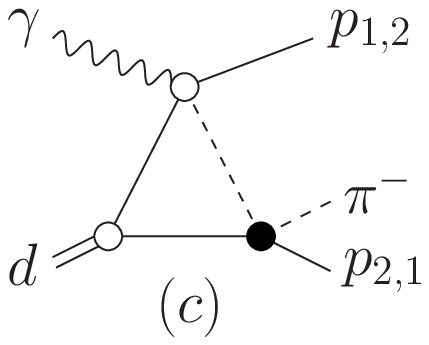}}

\caption{Feynman diagrams for the leading components of
        the $\rm \gamma d\rightarrow\pi^-pp$ amplitude.
        (a) Impulse approximation (IA), (b) pp-FSI, and
        (c) $\rm \pi N$-FSI. Filled black circles show
        FSI vertices. Wavy, dashed, solid, and double
        lines correspond to the photons, pions, nucleons,
        and deuterons, respectively.} \label{fig:fig3}
\end{figure}

Obviously, CLAS and A2 measurements are not enough to 
have compatible proton and neutron databases, 
specifically the energy binning of the CLAS 
measurements is 50~MeV or, in the worst case, 100~MeV 
while A2 Collaboration at MAMI measurements are able 
to have 2 to 4~MeV binning. 

\section{Neutron Data from Measurements with Deuteron Target}

\begin{figure}[htb!]
\centerline{
\includegraphics[scale=0.35,angle=0]{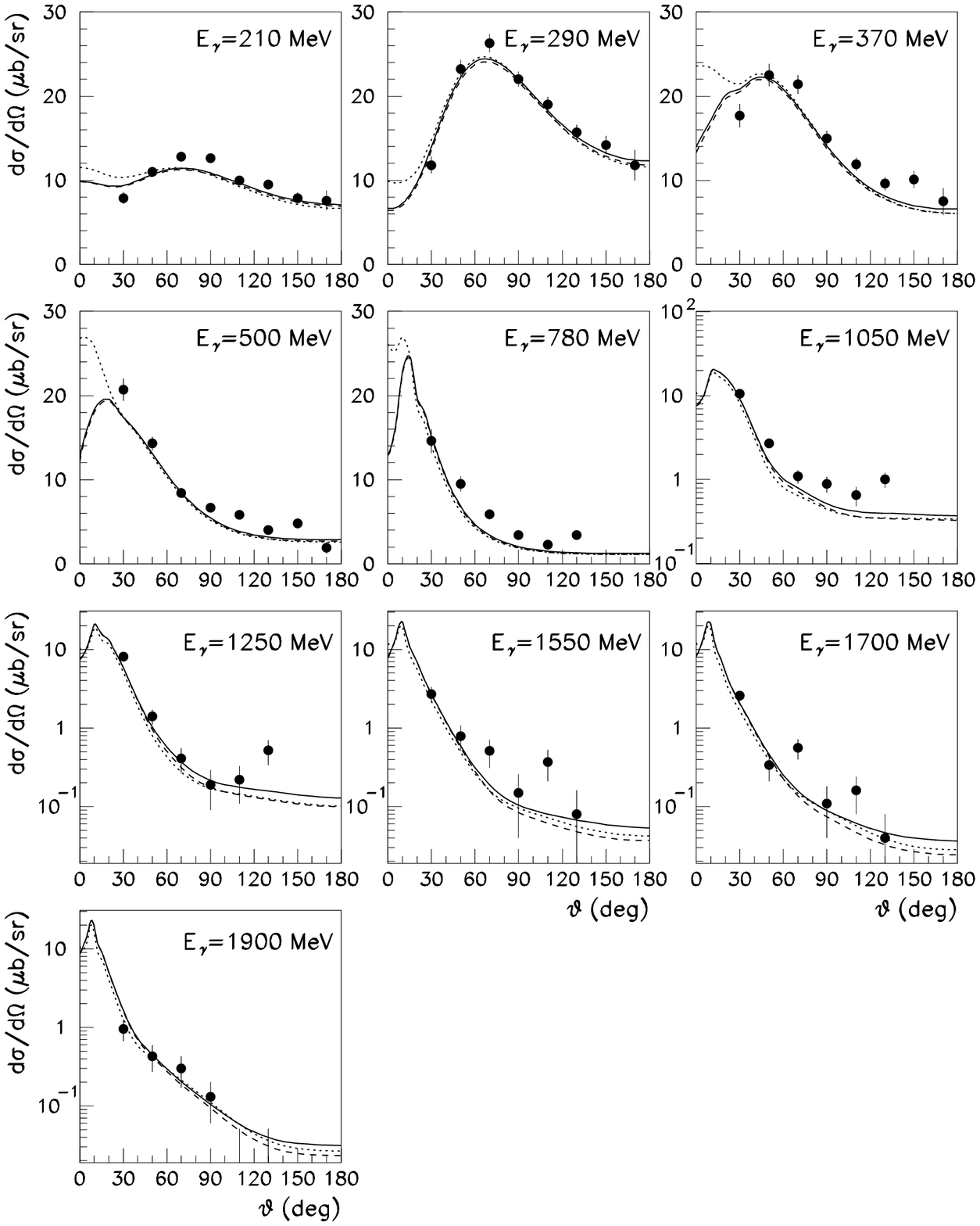}~~~~~~~~~
\includegraphics[scale=0.33,angle=0]{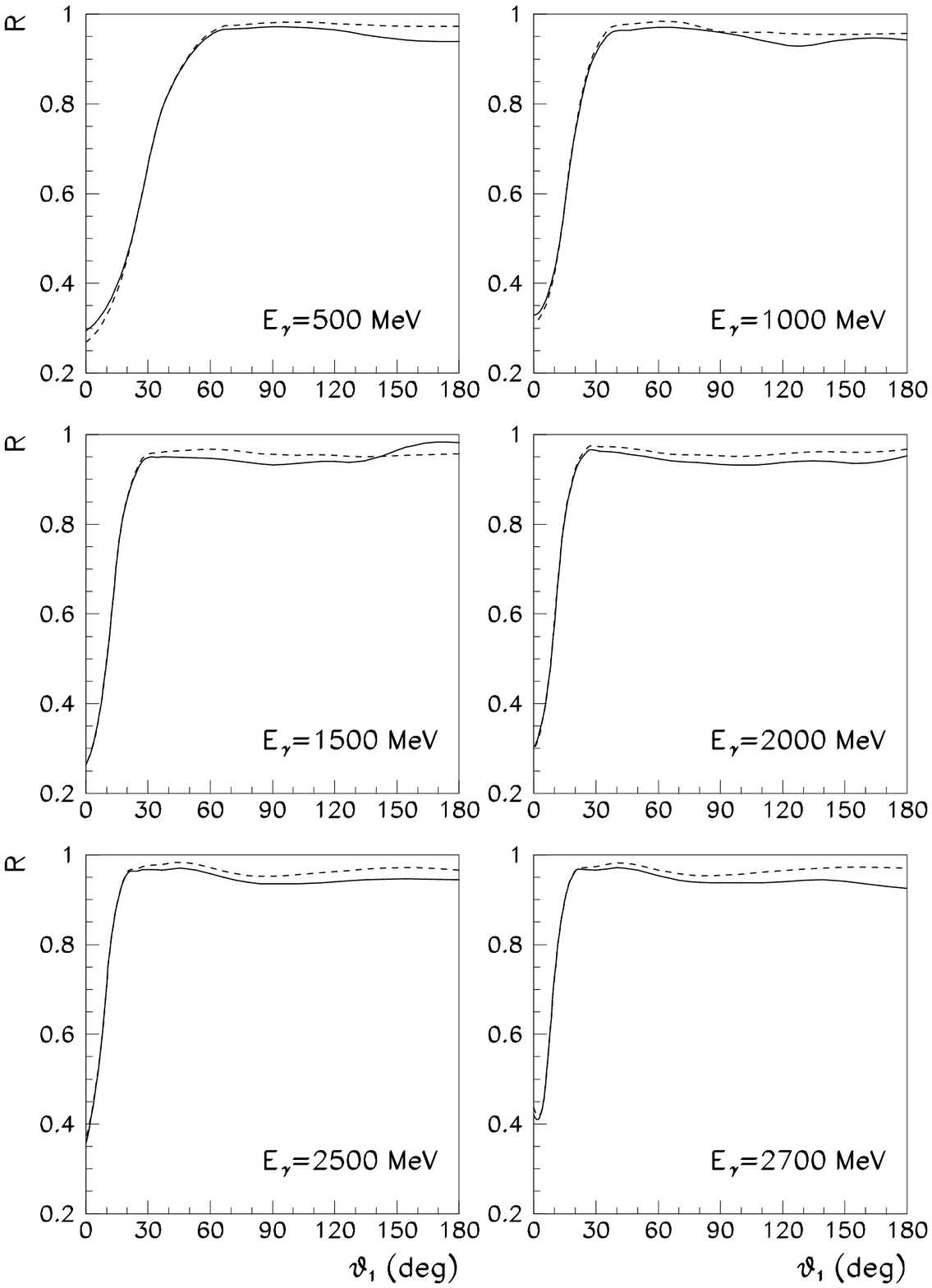}}

\caption{{\bf Left panel}: The differential cross
        section, $\rm d\sigma_{\rm \gamma d}/d\Omega$, of
        the reaction $\rm \gamma d\rightarrow\pi^-pp$
        in the laboratory frame at different values of
        the photon laboratory energy E $<$ 1900~MeV;
        $\rm \theta$ is the polar angle of the outgoing
        $\rm \pi^-$. Dotted curves show the contributions
        from the IA amplitude 
	[Fig.~\protect\ref{fig:fig3}(a)].
        Successive addition of the NN-FSI
        [Fig.~\protect\ref{fig:fig3}(b)] and $\rm \pi N$-FSI
        [Fig.~\protect\ref{fig:fig3}(c)] amplitudes 
	leads to dashed and solid curves, respectively. 
	The filled circles are the data from DESY bubble
        chamber~\protect\cite{ref19}.
        {\bf  Right panel}: The correction factor
        R(E,$\rm \theta$), where $\rm \theta$ is the polar angle
        of the outgoing $\pi^-$ in the rest frame of the
        pair $\rm \pi^-$+ fast proton. The kinematic cut,
        $\rm P_p >$ 200 MeV/c, is applied. The solid
        (dashed) curves are obtained with both $\rm \pi N$-
        and NN-FSI (only NN-FSI) taken into account.}
	\label{fig:fig4}
\end{figure}
The determination of the $\rm \gamma d\rightarrow
\pi^-pp$ differential cross sections with the FSI,
taken into account (including all leading diagrams
given in Fig.~\ref{fig:fig3}), were done recently
~\cite{ref8,ref14,ref15}, for the CLAS~\cite{ref14}
and MAMI~\cite{ref15} data. The SAID of GW Data
Analysis Center (DAC) phenomenological amplitudes
for $\rm \gamma N\rightarrow\pi N$~\cite{ref16},
$\rm NN \rightarrow NN$~\cite{ref17}, and $\rm \pi
N\rightarrow \pi N$~\cite{ref4} were used as inputs
to calculate the diagrams in Fig.~\ref{fig:fig3}.
The Bonn potential (full model)~\cite{ref18} was
used for the deuteron description. In Refs.
\cite{ref14,ref15}, we calculated the FSI correction
factor R(E,$\rm \theta$) dependent on photon energy, E,
and pion production angle in CM frame $\rm \theta$ and
fitted recent CLAS and MAMI $\rm d\sigma/d\Omega$ versus
the world $\rm \gamma N\rightarrow\pi N$ database
\cite{ref12} to get new neutron multipoles and
determine neutron resonance EM couplings~\cite{ref14}.

Results of calculations and comparison with the
experimental data on the differential cross sections,
$\rm d\sigma_{\rm \gamma d}/d\Omega$, where $\rm \Omega$ 
and $\rm \theta$ are solid and polar angles of outgoing
$\rm \pi^-$ in the laboratory frame, respectively, with
z-axis along the photon beam for the reaction $\rm
\gamma d \rightarrow\pi^-pp$ are given in
Fig.~\ref{fig:fig4} (left panels) for a number of the
photon energies, E.

The FSI corrections for the CLAS and MAMI quasi-free
kinematics were found to be small, as mentioned above.
As an illustration, Fig.~\ref{fig:fig4} (right panels)
shows the FSI correction factor $\rm R(E,\rm \theta)
=(d\sigma/d\Omega_{\pi p})/(d\sigma^{IA}/d\Omega_{\pi 
p})$ for the $\rm \gamma n\rightarrow\pi^-p$ 
differential cross sections as a function of the pion 
production angle in the CM ($\pi-p$) frame, $\theta$, 
for different energies over the range of the CLAS 
experiment. Overall, the FSI correction factor $\rm 
R(E,\rm \theta) < 1$, while the effect, i.e., the (1 
- R) value, vary from 10\% to 30\%, depending on the 
kinematics, and the behavior is very smooth versus pion 
production angle. We found a sizeable FSI-effect from 
S-wave part of pp-FSI at small angles. A small but 
systematic effect $|R - 1| << 1$ is found in the large 
angular region, where it can be estimated in the 
Glauber approach, except for narrow regions close to
$\rm \theta\sim 0^{\circ}$ or $\theta\sim 180^{\circ}$.  
The $\rm \gamma n\rightarrow\pi^0n$ case is much more 
complicated vs. $\rm \gamma n\rightarrow\pi^-p$ because 
in IA $\rm \pi^0n$ final state can come from both $\rm 
\gamma n$ and $\rm \gamma p$ initial interactions
\cite{ref20}. The leading diagrams for $\rm \gamma 
d\rightarrow\pi^0pn$ are similar as given on Fig.
\ref{fig:fig3}.

\section{New Neutron Amplitudes and Neutron EM Couplings}

The solution, SAID GB12~\cite{ref14}, uses the same 
fitting form as SAID recent SN11 solution~\cite{ref21}, 
which incorporated the neutron-target CLAS $\rm d\sigma / 
d\Omega$ for $\rm \gamma n\rightarrow\pi^-p$~\cite{ref14} 
and GRAAL $\Sigma$s for both $\rm \gamma n\rightarrow
\pi^-p$~\cite{ref22} and $\rm \gamma n\rightarrow\pi^0
n$~\cite{ref23}. This fit form was motivated by a 
multichannel K-matrix approach, with an added 
phenomenological term proportional to the $\pi N$ 
reaction cross section.
\begin{figure}[htb!]
{\centering
\includegraphics[scale=0.15,angle=90]{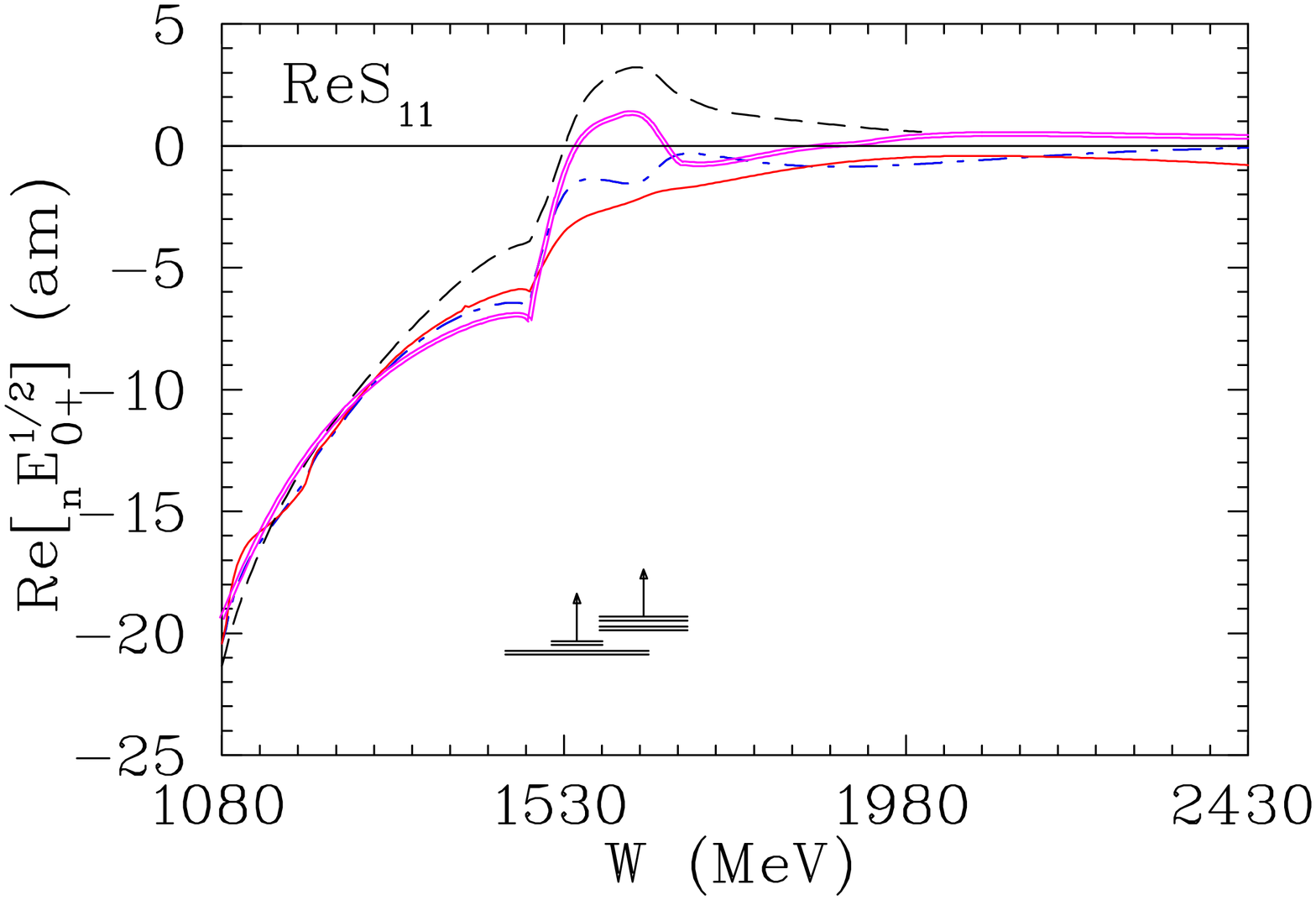}~~
\includegraphics[scale=0.15,angle=90]{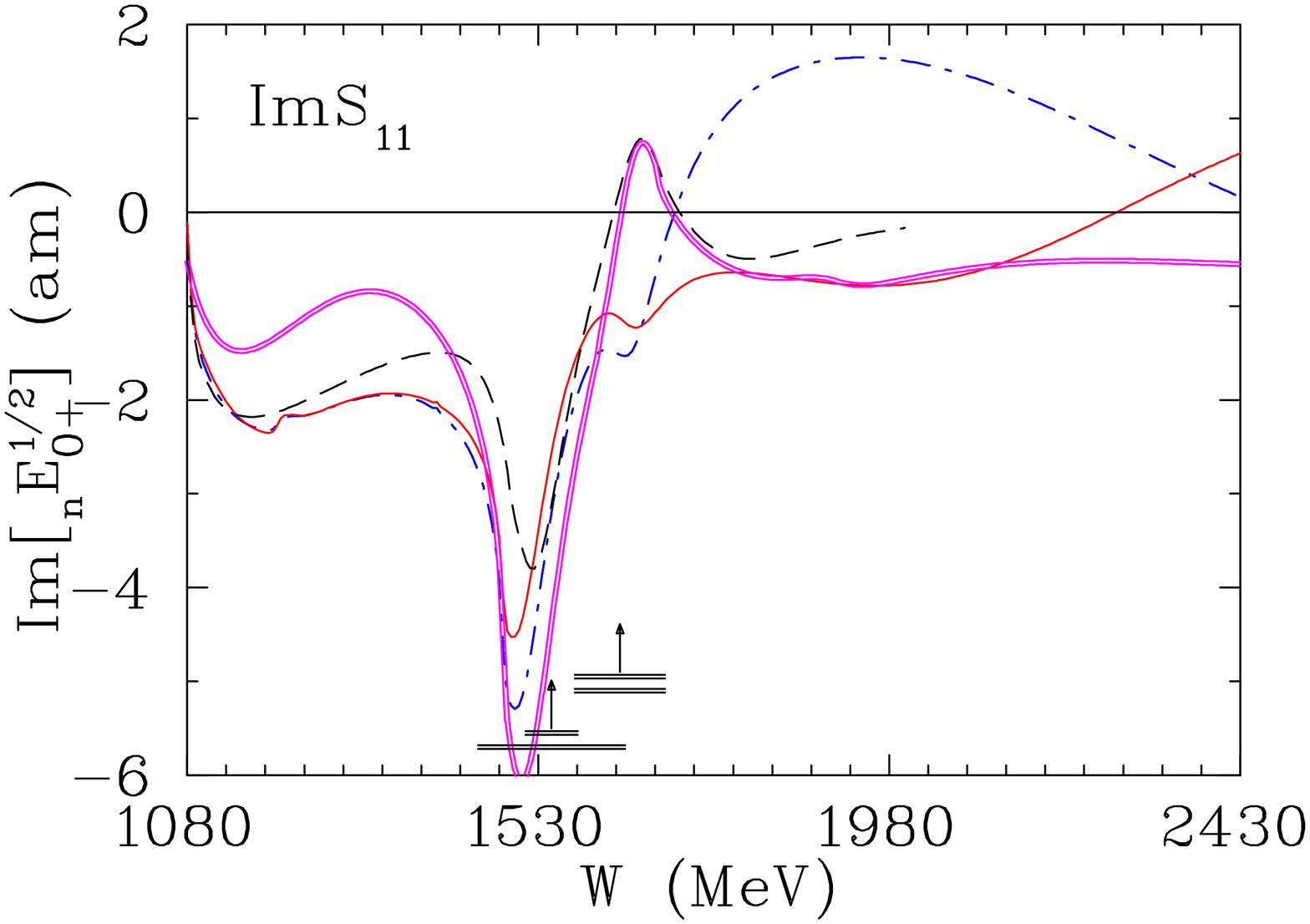}~~
\includegraphics[scale=0.15,angle=90]{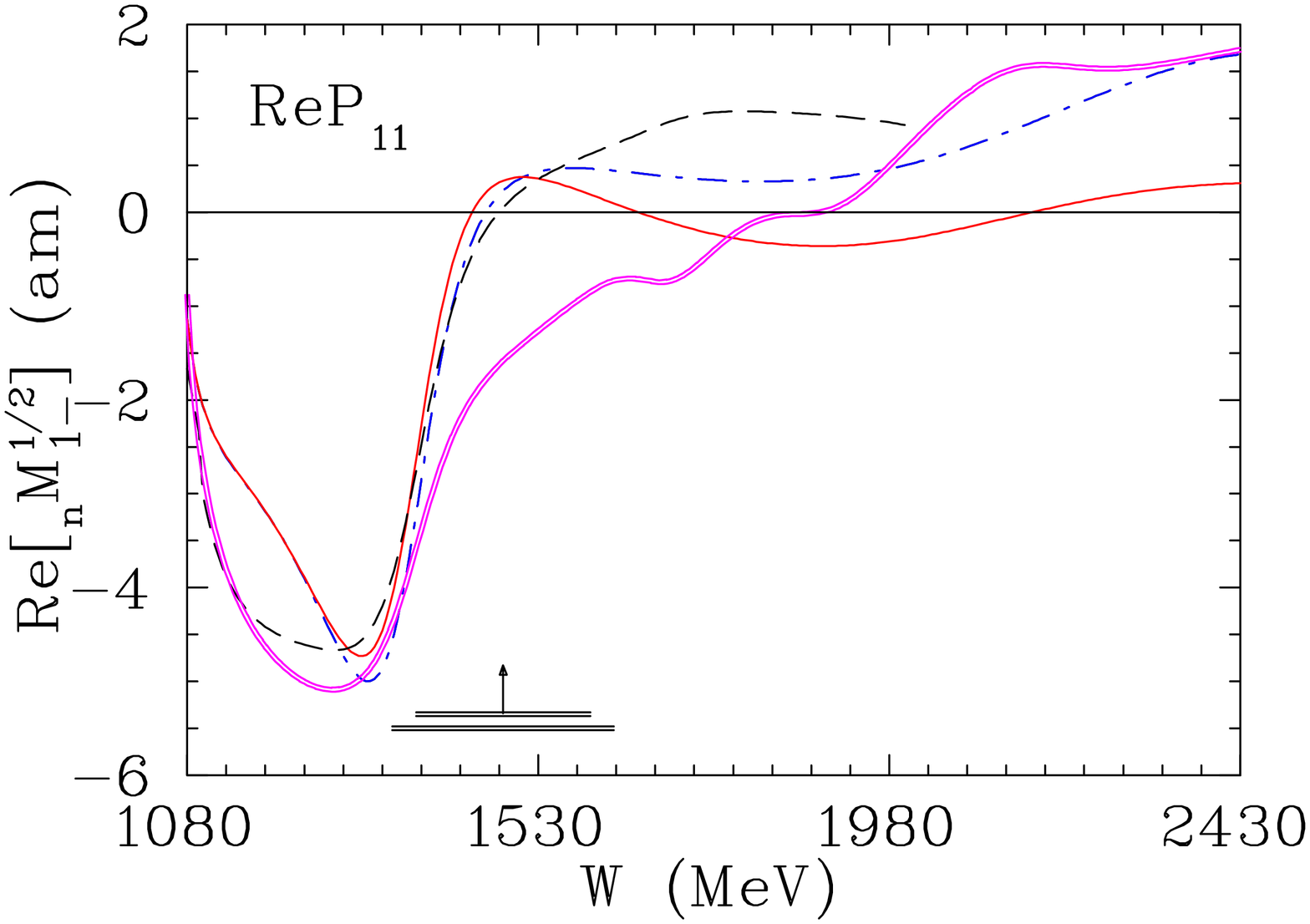}~~
\includegraphics[scale=0.15,angle=90]{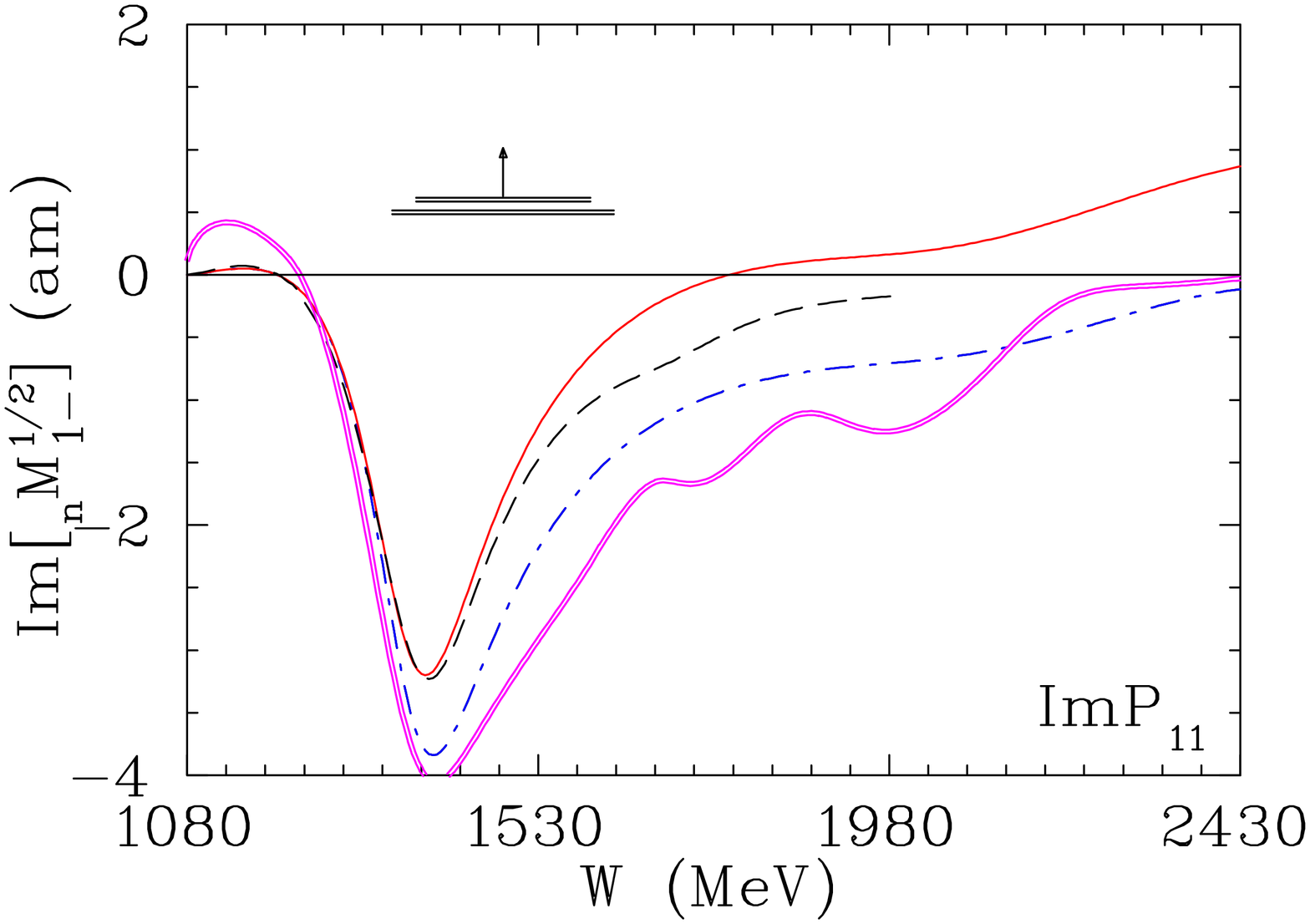}}

\caption{\small{Samples of neutron multipoles for I = 1/2. 
	Solid (dash-dotted) lines correspond to the SAID 
	GB12~\protect\cite{ref14} (SN11
        \protect\cite{ref21}) solution. Thick solid
        (dashed) lines give SAID GZ12~\protect\cite{ref14}
        solution (MAID07~\protect\cite{ref24}). Vertical
        arrows indicate mass (WR), and horizontal bars
        show full, $\rm \Gamma$, and partial, $\Gamma_{\rm
        \pi N}$, widths of resonances extracted by the
        Breit-Wigner fit of the ${\rm \pi N}$ data
        associated with the SAID solution WI08
        \protect\cite{ref4}.}} \label{fig:fig5}
\end{figure}
\begin{figure}[htb!]
{\centering
\includegraphics[scale=0.15,angle=90]{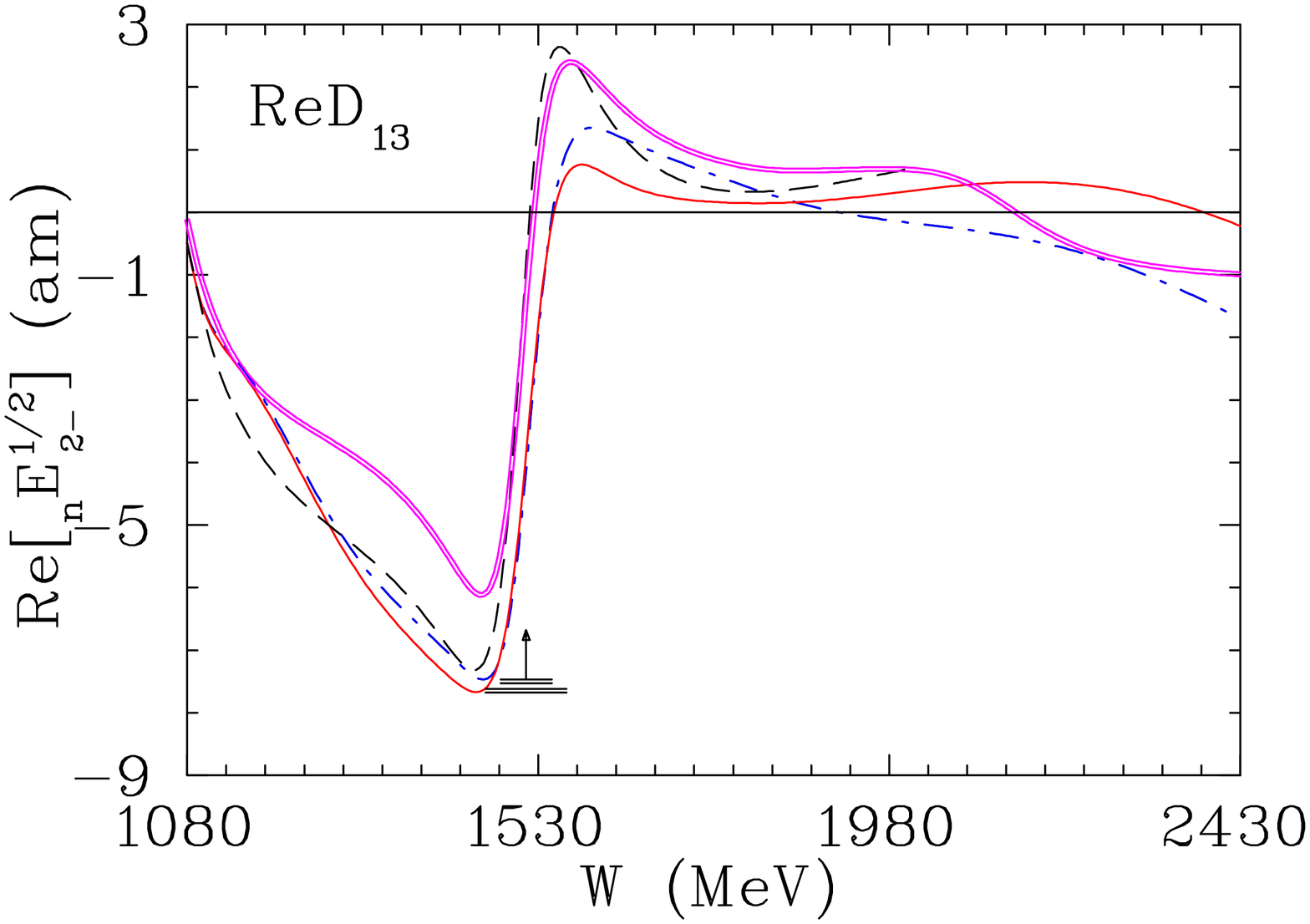}~~
\includegraphics[scale=0.15,angle=90]{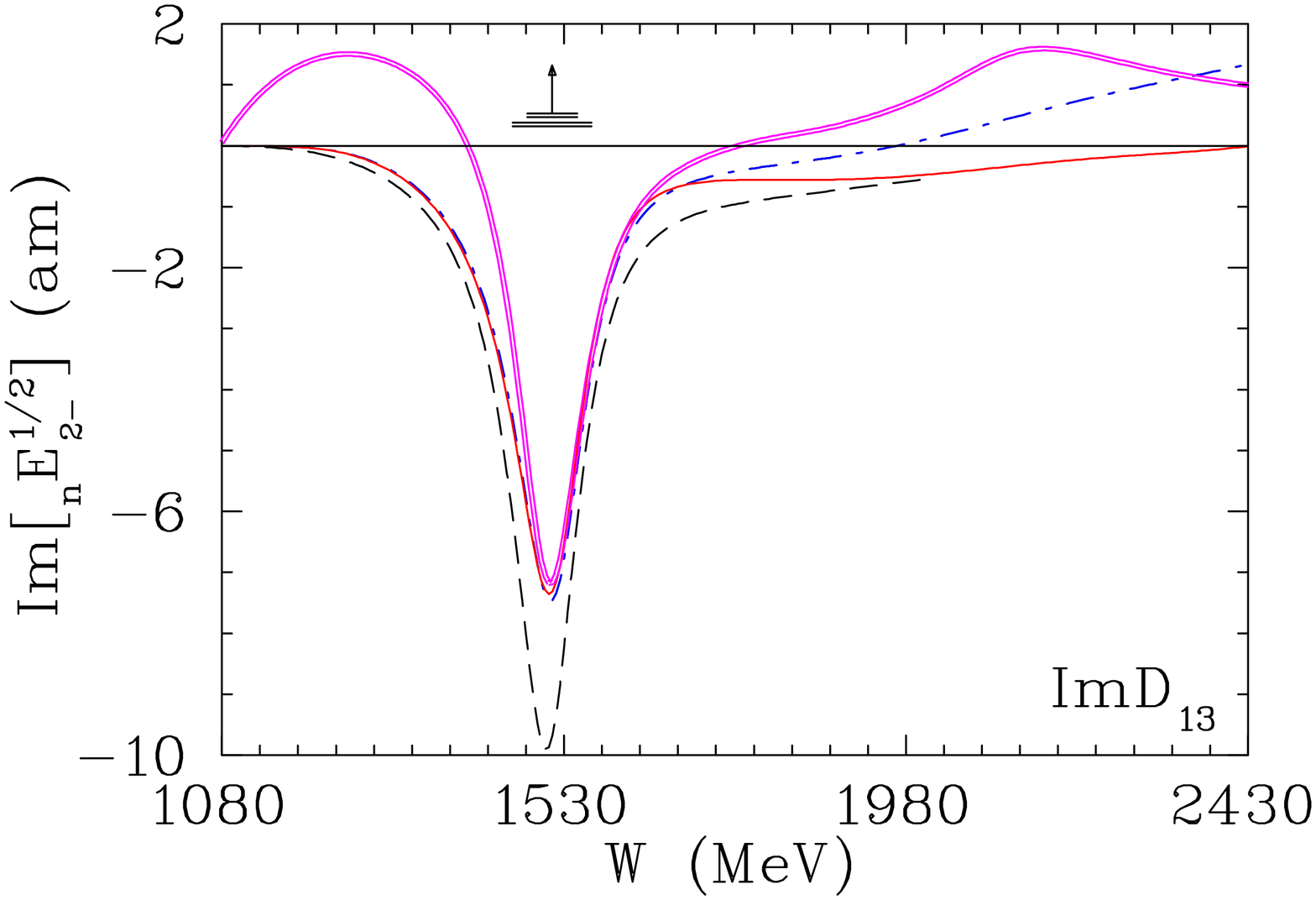}~~
\includegraphics[scale=0.15,angle=90]{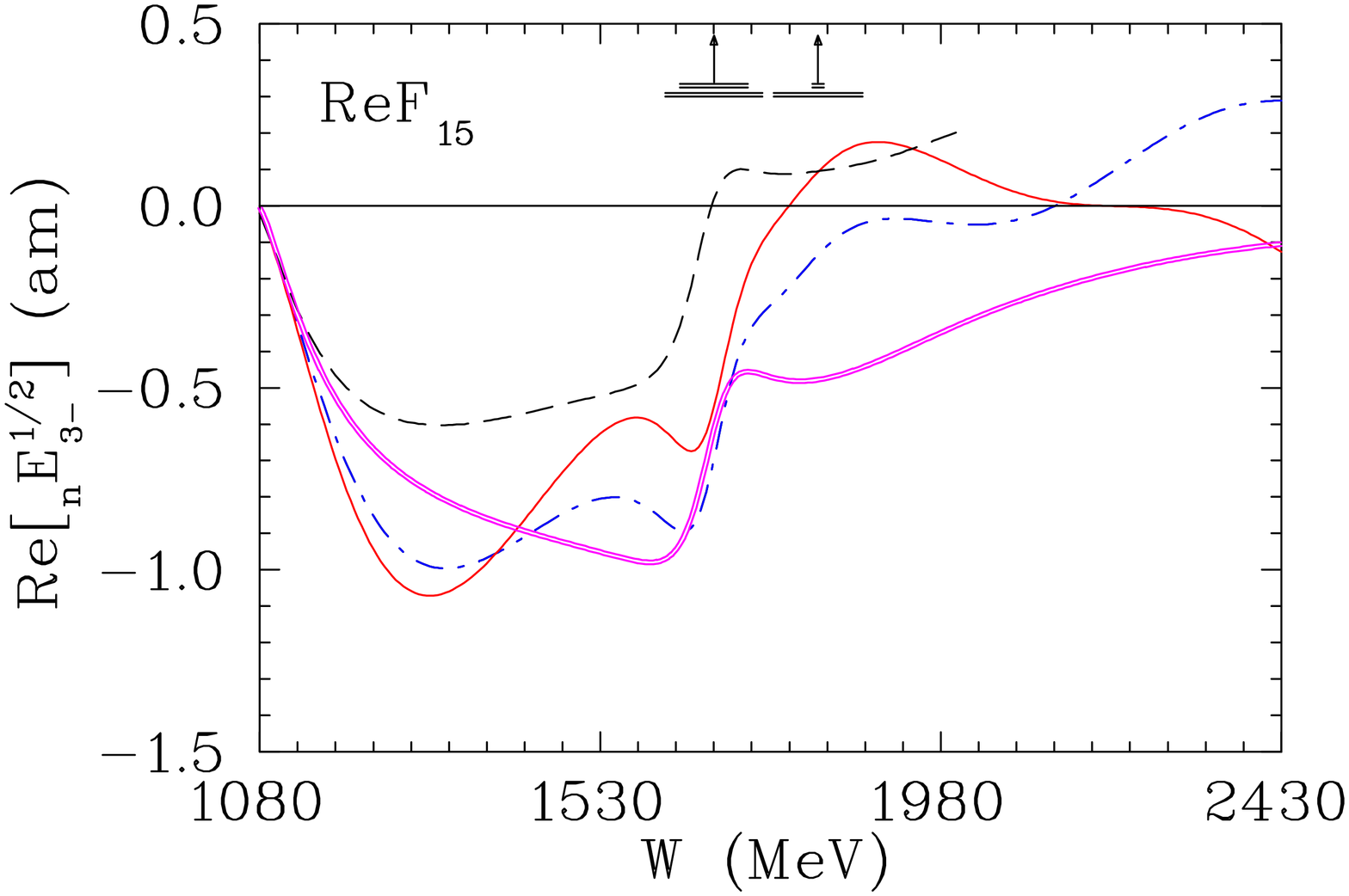}~~
\includegraphics[scale=0.15,angle=90]{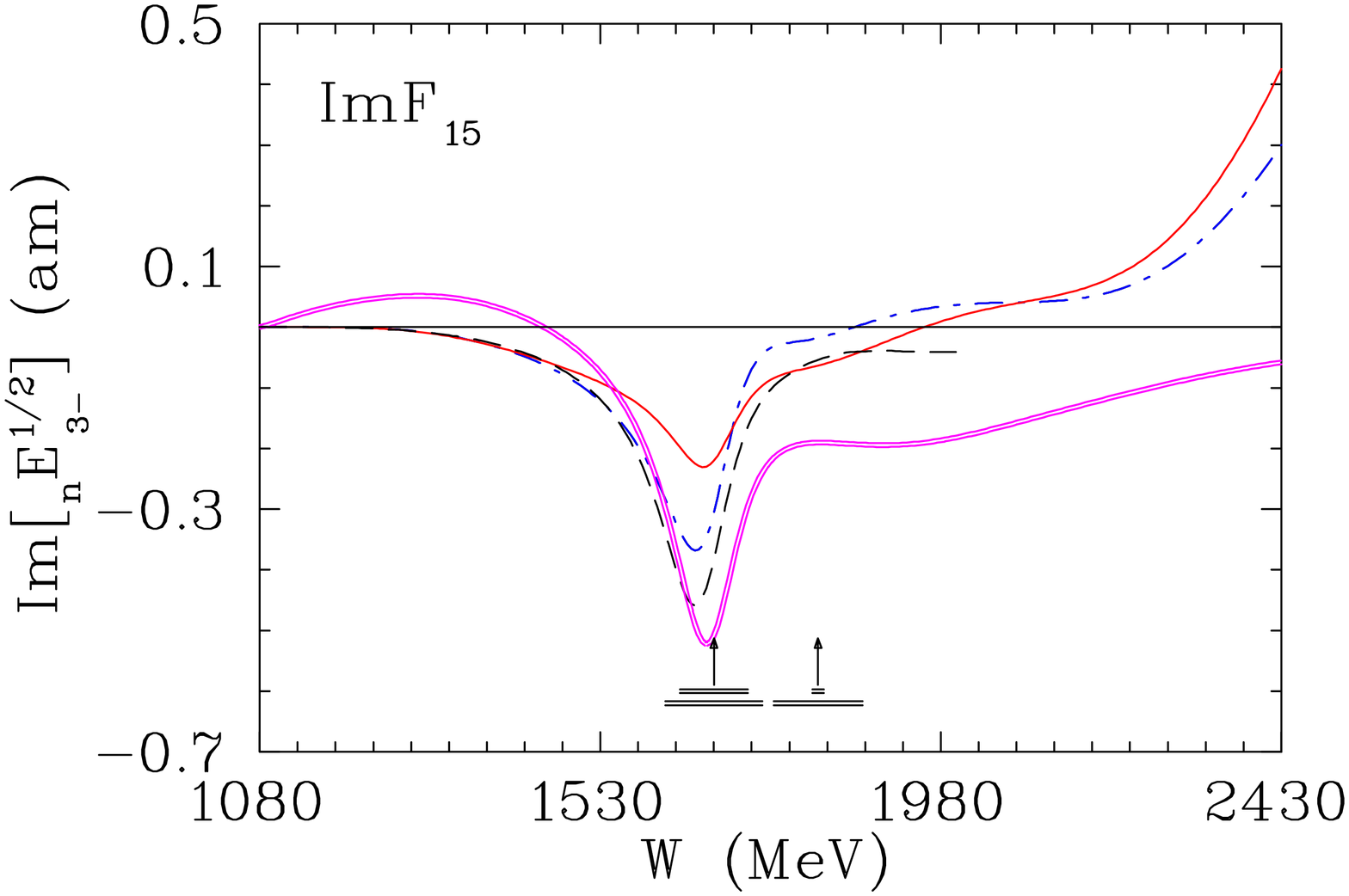}}

\caption{\small{Samples of neutron multipoles for I = 3/2. 
	Notation as in Fig.~\protect\ref{fig:fig5}}} 
	\label{fig:fig6}
\end{figure}
\begin{table}[htb!]
\centering
\caption{Neutron helicity amplitudes $\rm A_{1/2}$ and
        $\rm A_{3/2}$ (in [$\rm (GeV)^{-1/2}\times 10^{-3}$]
        units) from the SAID GB12~\protect\cite{ref14}
        (first row), previous SAID SN11~\protect\cite{ref21}
        (second row), recent BnGa13 by the Bonn-Gatchina
        group~\protect\cite{ref25} (third row), recent
        Kent12 by the Kent State Univ.
        group~\protect\cite{ref26} (forth row), and
        average values from the
        PDG14~\protect\cite{PDG} (fifth row).
	The relativized quark model predictions came from
	Ref.~\protect\cite{Ca92} (sixth row).}
\vspace{2mm}
{\begin{tabular}{|cc|ccc|c|}
\hline
Resonance      & nA$_{1/2}$ & Resonance      & nA$_{1/2}$ & nA$_{3/2}$  & Ref. \\
\hline
N(1535)1/2$^-$ & $-58\pm 6$ & N(1520)3/2$^-$ & $-46\pm 6$ & $-115\pm 5$ & SAID GB12\\
               & $-60\pm 3$ &                & $-47\pm 2$ & $-125\pm 2$ & SAID SN11\\
               & $-93\pm 11$&                & $-49\pm 8$ & $-113\pm 12$& BnGa13\\
               & $-49\pm 3$ &                & $-38\pm 3$ & $-101\pm 4$ & Kent12\\
               & $-46\pm 27$&                & $-59\pm 9$ & $-139\pm 11$& PDG14\\
               & $-63$      &                & $-38$      & $-114$      & Cap92\\
N(1650)1/2$^-$ & $-40\pm 10$& N(1675)5/2$^-$ & $-58\pm 2$ & $-80\pm 5$  & SAID GB12\\
               & $-26\pm 8$ &                & $-42\pm 2$ & $-60\pm 2$  & SAID SN11\\
               & $+25\pm 20$&                & $-60\pm 7$ & $-88\pm 10$ & BnGa13\\
               & $+11\pm 2$ &                & $-40\pm 4$ & $-68\pm 4$  & Kent12\\
               & $-15\pm 21$&                & $-43\pm 12$& $-58\pm 13$ & PDG14\\
               & $-35$      &                & $-35$      & $-51$       & Cap92\\
N(1440)1/2$^+$ & $+48\pm 4$ & N(1680)5/2$^+$ & $+26\pm 4$ & $-29\pm 2$  & SAID GB12\\
               & $+45\pm 15$&                & $+50\pm 4$ & $-47\pm 2$  & SAID SN11\\
               & $+43\pm 12$&                & $+34\pm 6$ & $-44\pm 9$  & BnGa13\\
               & $+40\pm 5$ &                & $+29\pm 2$ & $-59\pm 2$  & Kent12\\
               & $+40\pm 10$&                & $+29\pm 10$& $-33\pm 9$  & PDG14\\
               & $-6$       &                & $+19$      & $-23$       & Cap92\\
\hline
\end{tabular}} \label{tab:tab1}
\end{table}

However, these new CLAS cross sections departed
significantly from our predictions at the higher energies,
and greatly modified PWA result~\cite{ref14}
(Figs.~\ref{fig:fig5} and \ref{fig:fig6}). Following that, 
the BnGa group reported a neutron EM coupling 
determination~\cite{ref25} using the CLAS Collaboration 
$\rm d\sigma/d\Omega$ with our FSI~\cite{ref14} 
(Table~\ref{tab:tab1}). BnGa13~\cite{ref25} and SAID 
GB12~\cite{ref14} used the same (almost) data~\cite{ref14} 
to fit them while BnGa13 has several new Ad-hoc resonances
\cite{PDG}.

Overall: the difference between MAID07~\cite{ref24} with 
BnGa13 and SAID GB12 is rather small but resonances may be 
essentially different (Table~\ref{tab:tab1}). The new BnGa13 
has some difference vs. SAID GB12, PDG14~\cite{PDG}, for 
instance, for $\rm N(1535)1/2^-$, $\rm N(1650)1/2^-$, and 
$\rm N(1680)5/2^+$.

\section{Work in Progress}

At MAMI in March of 2013, we collected deuteron data below
E = 800~MeV with 4~MeV energy binning~\cite{ref27} and
will have a new experiment below E = 1600~MeV~\cite{ref28}
in the fall of 2016.

New 589 $\rm d\sigma/d\Omega$s by the A2 Collaboration at
MAMI contribution is about 160\% to the previous world
$\rm \pi^0n$ database.  Experiment is to have 19 angular
$\rm d\sigma/d\Omega$ for $-0.9 < cos\theta < +1$ and 31 
energy bins for E = 180 -- 800~MeV, $\Delta$E = 20~MeV (W 
= 1105 -- 1545~MeV, $\Delta$W = 10~MeV). Relative error of 
our measurement has a level of 1.5 -- 3\%.
\begin{figure}[htb!]
\centerline{
        \includegraphics[width=2.8in, height=1.3in, angle=0]{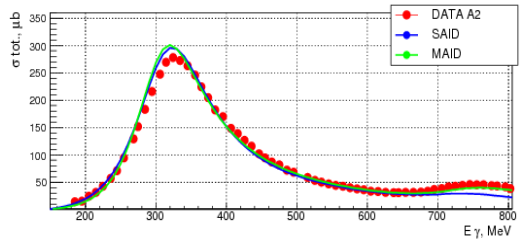}}

       \caption{Preliminary total cross sections for
        $\rm \gamma n\to\pi^0n$. See text for details.}
       \label{fig:fig7}
\end{figure}
Preliminary data for total cross section of the reaction
$\rm \gamma n\to\pi^0n$ shown on Fig.~\ref{fig:fig7}. 
This measurement is compared with a phenomenological 
result came from SAID~\cite{ref14} and MAID~\cite{ref24} 
PWAs. Comparison of our A2 Collaboration at MAMI 
experimental data with a MAID prediction gives more 
reasonable agreement, especially at higher energies.

Let us stress that the FSI corrections for the $\rm \pi^0$ 
photoproduction cross sections off the protons and neutrons 
are not equal in a general case. However, in a special case 
of the $\rm \Delta(1232)3/2^+$ energy region, the FSI 
corrections for $\rm \gamma p\to\pi^0p$ and $\rm \gamma n
\to\pi^0n$ cross sections are equal due to the isospin 
structure of the $\rm \gamma N\to\pi N$ amplitude~\cite{ref20}.

We are going to use our FSI technology to apply for the 
upcoming JLab CLAS (g13 run period) $d\sigma / d\Omega$ for 
$\rm \gamma n\rightarrow\pi^-p$ covering E = 400 - 2500~MeV 
and $\rm \theta = 18^{\circ} - 152^{\circ}$~\cite{ref30}. 
This data set will bring about 11k new measurements which 
quadruple the world $\rm \gamma n \rightarrow\pi^-p$ 
database. The ELPH facility at Tohoku Univ. will bring new 
$\rm d\sigma / d\Omega$ for $\rm \gamma n\rightarrow\pi^0n$ 
below E = 1200~MeV~\cite{ref31}.

\section{Conclusion}

\begin{itemize}
\item The differential cross section for the processes 
	$\rm \gamma n \rightarrow\pi^-p$ was extracted 
	from new CLAS and MAMI-B measurements accounting 
	for Fermi motion effects in the IA as well as 
	NN- and $\rm \pi N$-FSI effects beyond the IA.
\item Consequential calculations of the FSI corrections, 
	as developed by the GW-ITEP Collaboration, was 
	applied.
\item New cross sections departed significantly from our 
	predictions, at the higher energies, and greatly 
	modified the fit result.
\item New $\rm \gamma n\rightarrow\pi^-p$ and $\rm \gamma 
	n\rightarrow\pi^0n$ data will provide a critical 
	constraint on the determination of the multipoles 
	and EM couplings of low-lying baryon resonances 
	using the PWA and coupled channel techniques.
\item Polarized measurements at JLab/JLab12, MAMI, 
	SPring-8, CBELSA, and ELPH will help to bring 
	more physics in.
\item FSI corrections need to evaluate.
\end{itemize}

\section{Acknowledgments}

This material is based upon work supported by the U.S. 
Department of Energy, Office of Science, Office of 
Nuclear Physics, under Award Number DE-FG02-99-ER41110.



\begin{thebibliography}{1}
\bibitem{PDG} K.A.~Olive \textit{et al.} (Particle Data Group 
	Collaboration), Chin.\ Phys.\ C\ \textbf{38}, 090001 
	(2014).
\bibitem{ref2} K.M.~Watson, Phys.\ Rev.\ {\bf 95}, 228 (1954);
        R.L.~Walker, Phys.\ Rev.\ {\bf 182}, 1729 (1969).
\bibitem{ref4} R.A.~Arndt, W.J.~Briscoe, I.I.~Strakovsky, and
        R.L.~Workman, Phys.\ Rev.\ C\ {\bf 74}, 045205 (2006).
\bibitem{ref5} Ya.I.~Azimov and I.I.~Strakovsky, Proceedings 
	of the \textit{XVth International Conference on Hadron 
	Spectroscopy} (Hadron 2013), Nara, Japan, Nov. 2013, 
	PoS (Hadron 2014) 034.
\bibitem{ref6} A.B.~Migdal, JETP {\bf 1}, 2 (1955);
        K.M.~Watson, Phys.\ Rev.\ {\bf 88}, 1163 (1952).
\bibitem{ref8} V.E.~Tarasov, W.J.~Briscoe, H.~Gao, 
	A.E.~Kudryavtsev, and I.I.~Strakovsky, Phys.\ Rev.\ C\ 
	{\bf 84}, 035203 (2011).
\bibitem{hati} J.~Hadamard, \textit{Sur les problemes aux derivees
        partielles et leur signification physique}, (Princeton Univ.
        Bulletin. p.~49, 1902);
        A.N.~Tikhonov and V.Y.~Arsenin, \textit{Solutions of Ill-Posed
        Problems}, (New York: Winston, 1977).
\bibitem{ref10} A.M.~Sandorfi \textit{et al.}, 
	AIP\ Conf.\ Proc.\ {\bf 1432}, 219 
	(2012); K.~Nakayama, private communication, 2014.
\bibitem{ref12} W.J.~Briscoe, I.I.~Strakovsky, and 
	R.L.~Workman, Institute of Nuclear Studies of The 
	George Washington University Database;
        http://gwdac.phys.gwu.edu/analysis/pr\_analysis.html .
\bibitem{ref13} A.~Shafi \textit{et al.}, Phys.\ Rev.\ C\ 
	{\bf 70}, 035204 (2004).
\bibitem{ref14} W.~Chen \textit{et al.}, 
	Phys.\ Rev.\ C\ {\bf 86}, 015206 (2012).
\bibitem{ref15} W.J.~Briscoe \textit{et al.}, 
	Rev.\ C\ {\bf 86}, 065207 (2012).
\bibitem{ref16} M.~Dugger \textit{et al.}
	(CLAS Collaboration), 
	Phys.\ Rev.\ C\ {\bf 76}, 025211 (2007).
\bibitem{ref17} R.A.~Arndt, W.J.~Briscoe, I.I.~Strakovsky, and 
	R.L.~Workman, Phys.\ Rev.\ C\ {\bf 76}, 025209 (2007).
\bibitem{ref18} R.~Machleidt, K.~Holinde, and C.~Elster, Phys.\ 
	Rep.\ {\bf 149}, 1 (1987).
\bibitem{ref19} P.~Benz \textit{et al.}
        (Aachen-Bonn-Hamburg-Heidelberg-Muenchen Collaboration), 
	Nucl.\ Phys.\ B\ {\bf 65}, 158 (1973).
\bibitem{ref20} V.~Tarasov \textit{et al.}, 
	to be published in 
	Phys.\ At.\ Nucl.\ \textbf{79} (2016) ; 
	arXiv:1503.06671 [hep--ph].
\bibitem{ref21} R.L.~Workman \textit{et al.}, 
	Phys.\ Rev.\ C\ {\bf 85}, 025201 (2012).
\bibitem{ref22} G.~Mandaglio \textit{et al.} (GRAAL Collaboration), 
	Phys.\ Rev.\ C\ {\bf 82}, 045209 (2010).
\bibitem{ref23} R.~Di~Salvo \textit{et al.} (GRAAL Collaboration), 
	Eur.\ Phys.\ J.\ A\ {\bf 42}, 151 (2009).
\bibitem{ref24} D.~Drechsel, S.S.~Kamalov, and L.~Tiator, Eur.\ 
	Phys.\ J.\ A\ \textbf{34}, 69 (2007).
\bibitem{ref25} A.~Anisovich \textit{et al.},  
	Eur.\ Phys.\ J.\ A\ {\bf 49}, 67 (2013).
\bibitem{ref26} M.~Shrestha and D.M.~Manley, Phys.\ Rev.\ C\ 
	{\bf 86}, 055203 (2012).
\bibitem{Ca92} S.~Capstick, Phys.\ Rev.\ D\ \textbf{46}, 2864 
	(1992).
\bibitem{ref27} \textit{Meson production off the deuteron. I},
        Spokespersons: W.J.~Briscoe and I.I.~Strakovsky (A2 
	Collaboration at MAMI), MAMI Proposal MAMI--A2--02/12, 
	Mainz, Germany, 2012.
\bibitem{ref28} \textit{Meson production off the deuteron. II},
        Spokespersons: W.J.~Briscoe, V.V.~Kulikov, K.~Livingston, 
	and I.I.~Strakovsky (A2 Collaboration at MAMI), MAMI 
	Proposal MAMI--A2--02/13, Mainz, Germany, 2013.
\bibitem{ref30} P.~Mattione, Proceedings of the \textit{XVth 
	International Conference on Hadron Spectroscopy} 
	(Hadron 2013), Nara, Japan, Nov. 2013, PoS (Hadron 2014) 
	096.
\bibitem{ref31} T.~Ishikawa \textit{et al.}, Proceedings of the 
	\textit{XVth International Conference on Hadron 
	Spectroscopy} (Hadron 2013), Nara, Japan, Nov. 2013, 
	PoS (Hadron 2014) 095.
\end{thebibliography}
\end{document}